\documentclass[journal]{IEEEtran}

\usepackage[usenames]{color}

\usepackage{hyperref}
\usepackage{balance}
\usepackage{amsmath}
\usepackage{amssymb}
\usepackage{multirow}

\usepackage{cite}

\usepackage{ifpdf}

\ifCLASSINFOpdf
   \usepackage[pdftex]{graphicx}
   \graphicspath{{./pdf/}}
   \DeclareGraphicsExtensions{{.pdf}}
\else
   \usepackage[dvips]{graphicx}
   \graphicspath{{./eps/}}
   \DeclareGraphicsExtensions{{.eps}}
\fi

\begin{document}

\title{{The Outage Probability of a Finite Ad Hoc Network in Nakagami Fading}
\author{Don Torrieri and~Matthew C. Valenti}
\thanks{Paper approved by R. Schober, the Editor for Modulation and Signal Design of the IEEE Communications Society. Manuscript received Aug. 12, 2011; revised Mar. 26, 2012; accepted June 28, 2012.}
\thanks{M.C. Valenti's contribution was sponsored by the National Science Foundation under Award No. CNS-0750821 and by the United States Army Research Laboratory under Contract W911NF-10-0109.}
\thanks{D. ~Torrieri is with the US Army Research Laboratory, Adelphi, MD (email: dtorr@arl.army.mil).}
\thanks{M.~C.~Valenti is with West Virginia University, Morgantown, WV, U.S.A. (email: valenti@ieee.org).}%
\thanks{Digital Object Identifier 10.1109/TCOMM.2012.01.XXXXXX}}
\maketitle

\markboth{IEEE TRANSACTIONS ON COMMUNICATIONS, ACCEPTED FOR PUBLICATION}
{TORRIERI and VALENTI: The Outage Probability of a Finite Ad Hoc Network in Nakagami Fading}

\begin{abstract}
An ad hoc network with a finite spatial extent and number of nodes or
mobiles is analyzed. The mobile locations may be drawn from any spatial
distribution, and interference-avoidance protocols or protection against
physical collisions among the mobiles may be modeled by placing an exclusion
zone around each radio. The channel model accounts for the path loss,
Nakagami fading, and shadowing of each received signal. The Nakagami
m-parameter can vary among the mobiles, taking any positive value for each
of the interference signals and any positive integer value for the desired
signal. The analysis is governed by a new exact expression for the outage
probability, defined to be the probability that the
signal-to-interference-and-noise ratio (SINR) drops below a threshold, and
is conditioned on the network geometry and shadowing factors, which have
dynamics over much slower timescales than the fading. By averaging over many
network and shadowing realizations, the average outage probability and
transmission capacity are computed. 
Using the analysis, many aspects of the network performance are illuminated.
For example, one can determine the influence
of the choice of spreading factors, the
effect of the receiver location within the finite network region, and the
impact of both the fading parameters and the attenuation power laws.
\end{abstract}

\hyphenation{multi-symbol}

\thispagestyle{empty}


\section{Introduction}

\label{Section:Intro} An \textit{ad hoc network} or peer-to-peer network
comprises autonomous nodes or mobiles that communicate without a centralized
control or assistance. Ad hoc networks, which have both commercial and
military applications, possess no supporting infrastructure, fixed or
mobile. In addition to being essential when a cellular infrastructure is not
possible, ad hoc networks provide more robustness and flexibility in the
presence of node or mobile failures than cellular networks. In this paper,
we evaluate the performance of ad hoc networks of independent, identical
mobiles that can operate simultaneously in the same spectral band.

In the quest for analytical tractability, most authors (e.g., \cite{weber} - 
\cite{fran} and the many references therein) assume the ad hoc network has
an infinite number of mobiles spread over an infinite area and use the
methods of stochastic geometry \cite{stoy}. The locations of the mobiles at
any time instant are assumed to be spatially distributed as a Poisson point
process, and performance measures are derived by averaging over the spatial
distribution. When the network is finite, the Poisson point process is an
inadequate model primarily because it allows an unbounded number of mobiles
and does not account for the effects of the network boundary. However, this
model is extremely useful because it enables authors to use Campbell's
theorem \cite{bacc1}, \cite{stoy} to average over the possible network
realizations and obtain tractable mathematical expressions that vastly
simplify the performance analysis. Attempts to escape from the limitations
of the Poisson point process \cite{card} have involved restrictive
assumptions such as a low density of mobiles \cite{ganti} or operation in
the high-reliability regime \cite{giac}. A universal assumption is identical
fading models for all interference and reference signals, for instance
Nakagami fading with a common Nakagami-m parameter.

The analysis presented in this paper discards all previous
assumptions about the spatial distribution of the network mobiles and
identical fading. In this paper, the spatial extent of the network and
number of mobiles are finite. Each mobile has an arbitrary location
distribution with an allowance for the mobile's duty factor, shadowing, and
possible exclusion zones. A new analysis is presented that computes the 
\emph{exact} outage probability\footnote[1]{Outage probability is defined to be the probability that the
signal-to-interference-and-noise ratio (SINR) drops below a threshold.} in
the presence of Nakagami fading conditioned on the network geometry and
shadowing. The Nakagami-m parameter can vary among the mobiles, and the
reference receiver need not be at the center of the network area.

The main result in this paper is a closed-form expression for the
conditional outage probability at a reference receiver, which is conditioned on the location of the
interferers and the realization of the shadowing. The expression averages
over the fading, which has timescales much faster than that of the shadowing
or node location. The channel from each mobile to the reference receiver may have its own distinct Nakagami-m
parameter, and the ability to vary the Nakagami-m parameters can be used to model differing line-of-sight conditions
between the reference receiver and each mobile. The outage
probability in the presence of Nakagami-m fading has been previously
considered for interference-limited systems in \cite{dayya} - \cite{alouini}%
. None of these previous results accommodate thermal noise, and none are in
closed form because they require the computation of derivatives \cite%
{dayya,zhang} or an infinite series \cite{tellambura,alouini} for the most
general case of interference with non-integer Nakagami-m parameters.
Furthermore, these previous works assume interference with fixed power and
therefore do not consider the impact of the spatial model.

If unconditional outage probabilities are desired, then the conditional
outage probability at the reference receiver 
can be averaged over the possible locations of the interfering mobiles and
over the shadowing realizations. In limited cases, the averaging can be done
analytically, and the Appendix derives the spatially averaged outage
probability for the case that the reference receiver is at the center of an
annular-shaped network, the interferers are uniformly distributed, and there
is no shadowing. The unconditional outage probability can be computed for
more general cases through the use of Monte Carlo simulation, which involves
taking the numerical average over many network realizations, with the
networks drawn according to the desired distributions that model the
interferer locations and the shadowing. Because the fading does not need to
be realized, such a simulation is extremely fast, with execution times that
are competitive with the time required to compute analytical expressions
when they are available.

A major advantage of our approach over others is that any spatial model can
be analyzed, and the analysis can consider more than mere averages over
interfering mobile locations. For instance, the probability that a network realization
with specific mobile locations leads to a reference-receiver performance that 
fails to meet an outage constraint (the \emph{%
network} outage probability) can be determined. Since our model allows
realistic deployments, it can be incorporated into time-dependent \ random
geometric graphs that capture the dynamic network status \cite{li}. At each
sampling instant, each mobile receiver has specified parameter values for
the attenuation power law, the fading, and the shadowing from each signal
source based on the terrain and mobile locations at that instant. The outage
probabilities can be calculated by applying the closed-form equation
presented in this paper. Our model alleviates many problems of other
analytical models such as inability to handle arbitrarily located mobiles,
simple link models with disc signal coverage, and no modeling of border
effects.

The remainder of the paper is as follows. Section \ref{Section:SystemModel}
presents a physical interference model of the network. Section \ref%
{Section:OutageProbability} presents the derivation and description of the
conditional outage probability. Some example outage-probability calculations
are given in Section \ref{Section:Examples}. The issues of spatial averaging
and shadowing are discussed in Section \ref{Section:SpatialAveraging}, while
Section \ref{Transmission Capacity} examines the effects of mobile density
on both the outage probability and the transmission capacity. Conclusions
are drawn in Section \ref{Section:Conclusions}, while the Appendix derives a
closed-form expression for the spatially averaged outage probability in the
absence of shadowing when the reference receiver is at the center of an
annular network of uniformly-distributed interferers.

\vspace{-0.5cm}
\section{Network Model}

\label{Section:SystemModel} The network comprises $M+2$ mobiles that include
a reference receiver, a desired or reference transmitter $X_{0}$, and $M$
interfering mobiles $X_{1},...,X_{M}.$ The coordinate system is selected
such that the reference receiver is at the origin. The variable $X_{i}$
represents both the $i^{th}$ mobile and its location, and $||X_{i}||$ is the
distance from the $i^{th}$ mobile to the receiver. The mobiles can be
located in any arbitrary two- or three-dimensional regions. Two-dimensional
coordinates are conveniently represented by allowing $X_{i}$ to assume a
complex value, where the real component is the East-West coordinate and the
imaginary component is the North-South coordinate.

Although the network area may have any arbitrary shape, we assume that it is
a circular region of radius $r_{net}$ in the subsequent examples. While the
reference receiver is always located at the origin, the circular region
encompassing the network does not need to be centered at the origin.  Allowing
the network to be centered at a coordinate other than the origin is equivalent
to allowing the reference receiver to be located away from the center of the network.
A circular \emph{exclusion}
zone of radius $r_{ex}$ surrounds the reference receiver, and no interferers
are permitted within the exclusion zone. The exclusion zone is a type of 
\emph{guard zone}, and a nonzero $r_{ex}$ may be used to model the effects
of interference-avoidance protocols \cite{hasan}, \cite{krunz}, \cite{alaw}
or a prohibited zone to prevent physical collisions. When the
reference receiver is located the center of the circular network region, the
interferers are confined to an annulus with inner radius $r_{ex}$ and outer radius $r_{net}$. 
For computational and analytical convenience, we normalize $r_{net}$ so
that $r_{net}=1$ (equivalently, the unit distance is defined to be $r_{net}$%
).

Mobile $X_{i}$ transmits a signal whose average received power in the
absence of shadowing is $P_{i}$ at a reference distance $d_{0}$. Typically,
a common $P_{i}=P_{0}$ is used for all mobiles since the unknown shadowing
conditions make it difficult to adjust the power on a per-link basis.
Signals may be spread with a direct-sequence waveform. Let $G$ denote the 
\emph{processing gain} or \emph{spreading factor} of the direct-sequence
spread-spectrum signal, which is equal to the increase in bandwidth ($G=1$
indicates no spreading).

In the ad hoc network, the multiple-access communications are asynchronous.
Since there is no significant advantage to using short sequences, long
spreading sequences are assumed and modeled as random binary sequences with
chip duration $T_{c}$. The processing gain $G$ directly reduces the
interference power. The power of a multiple-access interference signal is
further reduced by the chip factor $h(\tau_{o})$, which is a function of the
chip waveform and the timing offset $\tau_{o}$ of the interference spreading
sequence relative to that of the desired or reference signal. Since only
timing offsets modulo-$T_{c}$ are relevant, $0\leq\tau_{o}<T_{c}.$ In a
network of quadriphase direct-sequence systems, a multiple-access
interference signal with power $I$ before despreading is reduced after
despreading to the power level $Ih(\tau_{o})/G,$ where \cite{torr}, \cite%
{torr2} 
\begin{equation}
h(\tau_{o})=\frac{1}{T_{c}^{2}}\left[ R_{\psi}^{2}(\tau_{o})+R_{%
\psi}^{2}(T_{c}-\tau_{o})\right]
\end{equation}
and $R_{\psi}(\tau_{o})$ is the partial autocorrelation for the normalized
chip waveform. Thus, the interference power is effectively reduced by the
factor $G/h(\tau_{o}).$ A \textit{rectangular chip waveform} has $R_{\psi
}(\tau_{o})=\tau_{o},$ and hence%
\begin{equation}
h(\tau_{o})=1+2\left( \frac{\tau_{o}}{T_{c}}\right) ^{2}-2\left( \frac {%
\tau_{o}}{T_{c}}\right)
\end{equation}
which implies that $1/2\leq h(\tau_{o})\leq1.$ If $\tau_{o}$ is assumed to
have a uniform distribution over [0, $T_{c}],$ then the expected value of $%
h(\tau_{o})$ is 2/3. It is assumed henceforth that $G/h(\tau_{o})$ is a
constant equal to $G/h$ for all mobiles in the network. After despreading,
the power of $X_{i},i>0$ (i.e., $X_{i}$ is an interferer), in the absence of
shadowing at distance $d_{0}$ is $\tilde{P}_{i}=P_{i}(h/G)$. For $X_{0}$
(i.e., the reference transmitter), the power after despreading remains $%
\tilde{P}_{0}=P_{0}$.

After despreading, $X_{i}$'s signal is received at the reference receiver
with an instantaneous power 
\begin{equation}
\rho_{i}=\tilde{P}_{i}g_{i}10^{\xi_{i}/10}f\left( ||X_{i}||\right)
\label{eqn:power}
\end{equation}
where $g_{i}$ is the power gain due to fading, $\xi_{i}$ is a \textit{%
shadowing factor}, and $f(\cdot)$ is a path-loss function. The \{$g_{i}\}$
are independent identically distributed (i.i.d.) and unit-mean with $%
g_{i}=a_{i}^{2}$, where $a_{i}$ is Nakagami with parameter $m_{i}$. In
Rayleigh fading, $m_{i}=1$ for all $i$, and the \{$g_{i}\}$ are
exponentially distributed. In the presence of log-normal shadowing, the $%
\{\xi_{i}\}$ are i.i.d. zero-mean Gaussian with variance $\sigma_{s}^{2}$.
In the absence of shadowing, $\xi_{i}=0$. \ For $d\geq d_{0}$, the path-loss
function is expressed as the attenuation power law 
\begin{equation}
f\left( d\right) =\left( \frac{d}{d_{0}}\right) ^{-\alpha}
\label{eqn:pathloss}
\end{equation}
where $\alpha\geq2$ is the attenuation power-law exponent, and it is assumed
that $d_{0}$ is sufficiently large that the signals are in the far field.

It is assumed that the \{$g_{i}\}$ remain fixed for the duration of a time
interval, but vary independently from interval to interval (block fading).
With probability $p_{i}$, the $i^{th}$ interferer transmits in the same time
interval as the reference signal. The \emph{activity probability} $\{p_{i}\}$
can be used to model voice-activity factors or controlled silence. Although
the $\{p_{i}\}$ need not be the same, it is assumed that they are identical
in the subsequent examples.

The instantaneous SINR at the receiver is 
\begin{equation}
\gamma=\frac{\rho_{0}}{\displaystyle{\mathcal{N}}+\sum_{i=1}^{M}I_{i}\rho_{i}%
}  \label{eqn:SINR1}
\end{equation}
where $\mathcal{N}$ is the noise power and $I_{i}$ is is a Bernoulli
variable with probability $P[I_{i}=1]=p_{i}$ and $P[I_{i}=0]=1-p_{i}$.
Substituting (\ref{eqn:power}) and (\ref{eqn:pathloss}) into (\ref{eqn:SINR1}%
) yields 
\begin{align}
\gamma & =\frac{P_{0}g_{0}10^{\xi_{0}/10}||X_{0}||^{-\alpha}}{\displaystyle%
\frac{\mathcal{N}}{d_{0}^{\alpha}}+\sum_{i=1}^{M}I_{i}\tilde {P}%
_{i}g_{i}10^{\xi_{i}/10}||X_{i}||^{-\alpha}}  \notag \\
& =\frac{g_{0}\Omega_{0}}{\displaystyle\Gamma^{-1}+\sum_{i=1}^{M}I_{i}g_{i}%
\Omega_{i}}  \label{Equation:SINR2}
\end{align}
where $\Gamma=d_{0}^{\alpha}P_{0}/\mathcal{N}$ is the signal-to-noise ratio
(SNR) when the reference transmitter is at unit distance and fading and
shadowing are absent, and 
\begin{equation}
\Omega_{i}=%
\begin{cases}
10^{\xi_{0}/10}||X_{0}||^{-\alpha} & i=0 \\ 
\displaystyle\frac{hP_{i}}{GP_{0}}10^{\xi_{i}/10}||X_{i}||^{-\alpha} & i\geq1%
\end{cases}
\label{eqn:omega}
\end{equation}
is the normalized received power due to $X_{i}$.

\section{Outage Probability}

\label{Section:OutageProbability}

Let $\boldsymbol{\Omega}=[\Omega_{0},...,\Omega_{M}]$ represent the set of
normalized powers given by (\ref{eqn:omega}). An \emph{outage} occurs when
the SINR $\gamma$ falls below an SINR threshold $\beta$, where $\gamma$ is
given for the particular $\boldsymbol{\Omega}$ by (\ref{Equation:SINR2}). It
follows that the outage probability $\epsilon$ for the given $\boldsymbol{%
\Omega}$ is 
\begin{equation}
\epsilon=P\left[ \gamma \leq \beta \big|\boldsymbol{\Omega}\right].
\label{Equation:Outage0}
\end{equation}
Because it is conditioned on $\boldsymbol{\Omega}$, the outage probability
is conditioned on the network geometry and shadowing factors, which have
dynamics over timescales that are much slower than the fading.
When conditioned on the fading and interference, the channel may be assumed to
be AWGN, and the capacity $C(\gamma)$ of the link conditioned on $%
\boldsymbol{\Omega}$ is a function of $\gamma$. The theoretically ideal 
SINR threshold may be found by inverting the capacity function for a given transmission rate $R$,
i.e., $\beta = C^{-1}(R)$.

Substituting (\ref{Equation:SINR2}) into (\ref%
{Equation:Outage0}), and rearranging yields 
\begin{equation}
\epsilon=P\left[ \left.
\beta^{-1}g_{0}\Omega_{0}-\sum_{i=1}^{M}I_{i}g_{i}\Omega_{i}\leq\Gamma^{-1}%
\right\vert \boldsymbol{\Omega}\right] .
\end{equation}
By defining 
\begin{equation}
\mathsf{S}=\beta^{-1}g_{0}\Omega_{0},\text{ }\mathsf{Y}_{i}=I_{i}g_{i}%
\Omega_{i}
\end{equation}%
\begin{equation}
\mathsf{Z}=\mathsf{S}-\sum_{i=1}^{M}\mathsf{Y}_{i}
\end{equation}
the outage probability may be expressed as 
\begin{equation}
\epsilon=P\left[ \mathsf{Z}\leq\Gamma^{-1}\big|\boldsymbol{\Omega}\right]
=F_{\mathsf{Z}}\left( \Gamma^{-1}\big|\boldsymbol{\Omega}\right)
\end{equation}
which is the cumulative distribution function (cdf) of $\mathsf{Z}$
conditioned on $\boldsymbol{\Omega}$ and evaluated at $\Gamma^{-1}$.

Define $\bar{F}_{\mathsf{Z}}(z)=1-F_{\mathsf{Z}}$ to be the complementary
cdf of $\mathsf{Z}$. Conditioned on $\boldsymbol{\Omega,}$ the complementary
cdf of $\mathsf{Z}$ is 
\begin{align}
\bar{F}_{\mathsf{Z}}\left( z\big|\boldsymbol{\Omega}\right) & =P[\mathsf{Z}>z%
\big|\boldsymbol{\Omega}]=P\left[ \mathsf{S}>z+\sum_{i=1}^{M}\mathsf{Y}_{i}%
\big|\boldsymbol{\Omega}\right] \nonumber \\
& =\underset{\mathbb{R}^{M}}{\int...\int}\left[ \int_{z+\sum y_{i}}^{\infty
}f_{\mathsf{S}}(s)ds\right] f_{\mathbf{Y}}(\mathbf{y})d\mathbf{y}
\label{eqn1}
\end{align}
where $f_{\mathbf{Y}}(\mathbf{y})$ is the joint probability density function
(pdf) of the vector $(\mathsf{Y}_{1},...,\mathsf{Y}_{M}),$ the pdf of the
gamma-distributed $\mathsf{S}$ with Nakagami parameter $m_{0}$ is%
\begin{equation}
f_{\mathsf{S}}(s)=\frac{{\left( \frac{\beta m_{0}}{\Omega_{0}}\right) }%
^{m_{0}}}{(m_{0}-1)!}{s}^{m_{0}-1}
\exp \left\{ - \frac{\beta m_{0} s}{\Omega_{0}} \right\}
\end{equation}
for $s \geq 0$, 
and the outer integral is over $M$-dimensional space.

Successive integration by parts and the assumption that $m_{0}$ is a
positive integer yield the solution to the inner integral: 
\begin{eqnarray}
\int_{z+\sum y_{i}}^{\infty}f_{\mathsf{S}}(s)ds
& = &
\exp\left\{ -\frac{\beta
m_{0}}{\Omega_{0}}(z+\sum y_{i})\right\} \nonumber \\
& &
\hspace{-1cm}
\times \sum_{s=0}^{m_{0}-1}\frac{1}{s!}{%
\left[ \frac{\beta m_{0}}{\Omega_{0}}(z+\sum y_{i})\right] }^{s}\hspace{-0.15cm}.
\label{eqn2}
\end{eqnarray}
Defining $\beta_{0}=\beta m_{0}/\Omega_{0}$, and substituting (\ref{eqn2})
into (\ref{eqn1}) yields 
\begin{eqnarray}
\bar{F}_{\mathsf{Z}}\left( z\big|\boldsymbol{\Omega}\right)
& = &
e^{-\beta_{0}z}\sum_{s=0}^{m_{0}-1}\frac{{\left( \beta_{0}z\right) }^{s}}{s!%
} \nonumber \\
& &
\hspace{-2cm}
\times
\underset{\mathbb{R}^{M}}{\int...\int}e^{-\beta_{0}\sum y_{i}}{\left(
1+z^{-1}\sum_{i=1}^{M}y_{i}\right) }^{s}f_{\mathbf{Y}}(\mathbf{y})d\mathbf{y}%
.  \label{eqn3}
\end{eqnarray}
Since $s$ is a positive integer, the binomial theorem indicates that 
\begin{equation}
{\left( 1+z^{-1}\sum_{i=1}^{M}y_{i}\right) }^{s}=\sum_{t=0}^{s}\binom{s}{t}%
z^{-t}\left( \sum_{i=1}^{M}y_{i}\right) ^{t}\hspace{-0.2cm}.  \label{eqn4}
\end{equation}
A multinomial expansion yields 
\begin{equation}
\left( \sum_{i=1}^{M}y_{i}\right) ^{t}=t!\mathop{ \sum_{\ell_i \geq 0}}%
_{\sum_{i=0}^{M}\ell_{i}=t}\left( \prod_{i=1}^{M}\frac{y_{i}^{\ell_{i}}}{%
\ell_{i}!}\right) .  \label{eqn5}
\end{equation}
where the summation on the right-hand side is over all sets of indices that
sum to $t$. Substituting (\ref{eqn4}) and (\ref{eqn5}) into (\ref{eqn3}), we
obtain 
\begin{eqnarray}
\bar{F}_{\mathsf{Z}}\left( z\big|\boldsymbol{\Omega}\right)
& = &
e^{-\beta_{0}z}\sum_{s=0}^{m_{0}-1}\frac{{\left( \beta_{0}z\right) }^{s}}{s!%
}\sum _{t=0}^{s}\binom{s}{t}z^{-t}t! \nonumber \\
& &
\hspace{-2cm}
\times \hspace{-0.25cm}
\mathop{ \sum_{\ell_i \geq 0}}%
_{\sum_{i=0}^{M}\ell_{i}=t}\underset{\mathbb{R}^{M}}{\int...\int}\left(
\prod_{i=1}^{M}e^{-\beta_{0}y_{i}}\frac{y_{i}^{\ell_{i}}}{\ell_{i}!}\right)
f_{\mathbf{Y}}(\mathbf{y})d\mathbf{y}.
\end{eqnarray}
Using the fact that the $\{\mathsf Y_i\}$ are nonnegative and assuming that the interfering channels fade independently, 
\begin{eqnarray}
\bar{F}_{\mathsf{Z}}\left( z\big|\boldsymbol{\Omega}\right)
& = &
e^{-\beta_{0}z}\sum_{s=0}^{m_{0}-1}\frac{{\left( \beta_{0}z\right) }^{s}}{s!%
}\sum _{t=0}^{s}\binom{s}{t}z^{-t}t! \nonumber \\
& &
\hspace{-1cm}
\times \hspace{-0.25cm}
\mathop{ \sum_{\ell_i \geq 0}}%
_{\sum_{i=0}^{M}\ell_{i}=t}\prod_{i=1}^{M}\int_{0}^{\infty}\frac{y^{\ell_{i}}%
}{\ell_{i}!}e^{-\beta_{0}y}f_{\mathsf{Y}_{i}}(y)dy.  \label{eqn7}
\end{eqnarray}

Taking into account the Nakagami fading and the activity probability $p_{i}$%
, the pdf of $f_{\mathsf{Y}_{i}}(y)$ is 
\begin{eqnarray}
f_{\mathsf{Y}_{i}}(y)
& = &
(1-p_{i})\delta(y) \nonumber \\
& &
\hspace{-1cm}
+ \; p_{i}\left( \frac{m_{i}}{\Omega_{i}}%
\right) ^{m_{i}}\frac{1}{\Gamma(m_{i})}y^{m_{i}-1}e^{-ym_{i}/\Omega_{i}}u(y) 
\end{eqnarray}
where $u(y)$ is the unit-step function, and $\delta(y)$ is the Dirac delta
function. Substituting this pdf, the integral in (\ref{eqn7}) is 
\begin{eqnarray}
\int_{0}^{\infty}\frac{y^{\ell_{i}}}{\ell_{i}!}e^{-\beta_{0}y}f_{\mathsf{Y}%
_{i}}(y)dy
& = &
(1-p_{i})\delta_{\ell_{i}} \nonumber \\
& &
\hspace{-5cm}
+ \; \left( \frac{p_{i}\Gamma(%
\ell_{i}+m_{i})}{\ell_{i}!\Gamma(m_{i})}\right) \left( \frac{\Omega_{i}}{%
m_{i}}\right) ^{\ell_{i}}\left( \beta_{0}\frac{\Omega_{i}}{m_{i}}+1\right)
^{-(m_{i}+\ell_{i})}\hspace{-1.25cm}  \label{eqn9}
\end{eqnarray}
where $\delta_{\ell}$ is the Kronecker delta function, equal to 1 when $%
\ell=0$, and zero otherwise. Substituting (\ref{eqn9}) into (\ref{eqn7}) \
and using 
\begin{equation}
\binom{s}{t}\left( \frac{t!}{s!}\right) =\left( \frac{s!}{t!(s-t)!}\right)
\left( \frac{t!}{s!}\right) =\frac{1}{(s-t)!}
\end{equation}
gives
\begin{eqnarray}
\bar{F}_{\mathsf{Z}}\left( z\big|\boldsymbol{\Omega}\right)
& = &
e^{-\beta_{0}z}\sum_{s=0}^{m_{0}-1}{\left( \beta_{0}z\right) }%
^{s}\sum_{t=0}^{s}\frac{z^{-t}}{(s-t)!} \nonumber \\
& &
\hspace{-2.25cm}
\times \hspace{-0.35cm}
\mathop{ \sum_{\ell_i \geq 0}}%
_{\sum_{i=0}^{M}\ell_{i}=t}\prod_{i=1}^{M}\left[ (1-p_{i})\delta_{\ell_{i}}+%
\frac{p_{i}\Gamma (\ell_{i}+m_{i})\left( \frac{\Omega_{i}}{m_{i}}\right)
^{\ell_{i}}}{\ell _{i}!\Gamma(m_{i})\left( \beta_{0}\frac{\Omega_{i}}{m_{i}}%
+1\right) ^{(m_{i}+\ell_{i})}}\right]. \nonumber \\
\end{eqnarray}
This equation may be written as 
\begin{equation}
\bar{F}_{\mathsf{Z}}\left( z\big|\boldsymbol{\Omega}\right)
=e^{-\beta_{0}z}\sum_{s=0}^{m_{0}-1}{\left( \beta_{0}z\right) }%
^{s}\sum_{t=0}^{s}\frac{z^{-t}H_{t}(\boldsymbol{\Psi})}{(s-t)!}
\label{Equation:NakagamiConditional}
\end{equation}
where 
\begin{align}
\Psi_{i} & =\left( \beta_{0}\frac{\Omega_{i}}{m_{i}}+1\right) ^{-1}\hspace{%
-0.5cm},\hspace{1cm}\mbox{for $i=\{1,...,M\}$, }  \label{Equation:Psi} \\
H_{t}(\boldsymbol{\Psi}) & =\mathop{ \sum_{\ell_i \geq 0}}%
_{\sum_{i=0}^{M}\ell_{i}=t}\prod_{i=1}^{M}G_{\ell_{i}}(\Psi_{i}),
\label{Equation:Hfunc}
\end{align}%
\begin{equation}
G_{\ell}(\Psi_{i})=%
\begin{cases}
1-p_{i}(1-\Psi_{i}^{m_{i}}) & \mbox{for $\ell=0$} \\ 
\frac{p_{i}\Gamma(\ell+m_{i})}{\ell!\Gamma(m_{i})}\left( \frac{\Omega_{i}}{%
m_{i}}\right) ^{\ell}\Psi_{i}^{m_{i}+\ell} & \mbox{for $\ell>0$.}%
\end{cases}
\label{Equation:Gfunc}
\end{equation}

Equation (\ref{Equation:Hfunc}) may be efficiently computed as follows. For
each possible $t=\{0,..,m_{0}-1\}$, precompute a matrix $\mathcal{I}_{t}$
whose rows contain all sets of non-negative indices $\{\ell_{1},...,\ell
_{M}\}$ that sum to $t$. There will be 
\begin{equation}
\binom{t+M-1}{t}
\end{equation}
rows and $M$ columns in $\mathcal{I}_{t}$. The $\mathcal{I}_{t}$'s may be
reused anytime the same $M$ is considered. Compute a row vector $\boldsymbol{%
\Psi}$ containing the $\Psi_{i}$. For each possible $\ell=\{0,..,m_{0}-1\}$,
compute (\ref{Equation:Gfunc}) using $\boldsymbol{\Psi}$ and place the
resulting row into an $m_{0}\times M$ matrix $\mathbf{G}$. Each term of (\ref%
{Equation:Hfunc}) can be found by using the corresponding row from $\mathcal{%
I}_{t}$ as an index into $\mathbf{G}$. Taking the product along the length
of the resulting row vector gives the corresponding term of the summation.
More generally, the entire $\mathcal{I}_{t}$ matrix can be used to index $%
\mathbf{G}$. To be consistent with matrix-based languages, such as Matlab,
denote the result of the operation as $\mathbf{G}(\mathcal{I}_{t})$. Taking
the product along the rows of $\mathbf{G}(\mathcal{I}_{t})$ and then the sum
down the resulting column gives (\ref{Equation:Hfunc}).

\section{Examples}

\begin{figure}[t]
\centering
\includegraphics[width=3.4in]{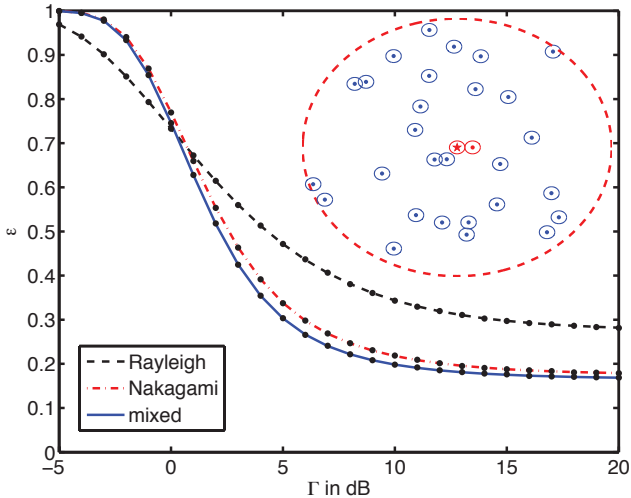}
\vspace{-0.2cm}
\caption{ An example network, which is drawn according to the uniform clustering model, is shown in the upper right portion of the figure. The reference receiver (indicated by the {\color{red} $\star $}) is placed at the origin, and the corresponding reference transmitter is located  to its right
(indicated by the {\color{red} $\bullet $} to the right of the receiver). $%
M=28$ interferers (indicated by the {\color{blue} $\bullet $}) are placed at random within the dashed circle. 
An exclusion zone surrounds each mobile (indicated by solid circles
surrounding the mobiles). The figure also shows the outage probability as
a function of SNR $\Gamma $, conditioned on the pictured network topology.
Performance is shown for three fading models without spreading or shadowing.
Analytical expressions are plotted by lines while dots represent simulation
results (with one million trials per point).}
\vspace{-0.4cm}
\label{Figure:FigA}
\end{figure}

\label{Section:Examples} 
Consider the network topology shown in the upper right corner of Fig. \ref{Figure:FigA}.
The reference receiver is at the center of the network, and $M=28$ interferers
are within an annular region with inner radius $r_{ex}=0.05$ and outer radius 
$r_{net}=1$. An
exclusion zone of radius $r_{ex}$ surrounds each mobile, and interfering
mobiles are uniformly distributed throughput the network area outside of the
exclusion zones. Mobiles are placed successively according to a \emph{%
uniform clustering} model as follows. Let $X_{i}=r_{i}e^{j\theta_{i}}$
represent the location of the $i^{th}$ mobile. A pair of independent random
variables $(y_{i},z_{i})$ is selected from the uniform distribution over $%
[0,1]$. From these variables, the location is initially selected according
to a uniform spatial distribution over a disk with radius $r_{net}$\ by
setting $r_{i}=\sqrt{y_{i}}r_{net}$ and $\theta_{i}=2\pi z_{i}$. If the
corresponding $X_{i}$ falls within an exclusion zone of one of the $i-1$
previous mobile locations, then a new random location is assigned to the $%
i^{th}$ mobile as many times as necessary until it falls outside any
exclusion zone.

Once the mobile locations $\{X_{i}\}$ are realized, the $\{\Omega_{i}\}$ are
determined by assuming an attenuation power-law exponent $\alpha=3.5$, a
common transmit power $P_{i}=P$ for all $i$, no shadowing, and that the
reference transmitter is at distance $||X_{0}||=0.1$ from the reference
receiver. For all results given in this paper, the value of $p_{i}=0.5$ for
all $i,$ and the SINR threshold is $\beta=0$ dB, which corresponds to the
unconstrained AWGN capacity limit for a rate-$1$ channel code and complex inputs.

\textbf{Example \#1.} Suppose that spread-spectrum modulation is not used ($%
G=h=1$) and that all signals undergo Rayleigh fading. Then, $m_{i}=1$ for
all $i$, $\beta_{0}=\beta/\Omega_{0}=\beta||X_{0}||^{\alpha}$, $\Omega_{i}=$ 
$||X_{i}||^{-\alpha},$ and (\ref{Equation:NakagamiConditional}) specializes
to 
\begin{equation}
\bar{F}_{\mathsf{Z}}\left( z\big|\boldsymbol{\Omega}\right)
=e^{-\beta_{0}z}\prod_{i=1}^{M}\frac{1+\beta_{0}(1-p_{i})\Omega_{i}}{%
1+\beta_{0}\Omega_{i}}.  \label{Equation:RayleighConditional}
\end{equation}
The outage probability can thus be readily found for any given realization
of $\boldsymbol{\Omega}$. The outage probability, found by evaluating (%
\ref{Equation:RayleighConditional}) at $z=\Gamma^{-1}$, is shown along
with the spatial locations of the mobiles in Fig. \ref{Figure:FigA}. Also
shown is the outage probability generated by simulation, which involves
randomly generating the mobile locations and the exponentially distributed
power gains $g_{0},...,g_{M}$. As can be seen in the figure, the analytical
and simulation results coincide, which is to be expected because (\ref%
{Equation:RayleighConditional}) is exact. Any discrepancy between the curves
can be attributed to the finite number of Monte Carlo trials (one million
trials were executed per SNR point).

\textbf{Example \#2.} Now suppose that the link between the source and
receiver undergoes Nakagami fading with parameter $m_{0}=4$. 
The outage probability, found using (\ref%
{Equation:NakagamiConditional}) through (\ref{Equation:Gfunc}), is also 
plotted in Fig. \ref{Figure:FigA}. The figure shows two choices for the
Nakagami parameter of the interferers: $m_{i}=1$
and $m_{i}=4,$ $i=1,2,\ldots,M$. The $m_{i}=4$ case, denoted by ``Nakagami'' in the figure legend,
 represents the situation
where the reference transmitter and interferers are equally visible to the
receiver.  The $m_{i}=1$ case, denoted by ``mixed'' in the figure legend, represents a more typical situation where
the interferers are not in the line-of-sight. As with the previous example,
the analytical curves are verified by simulations involving one million
Monte Carlo trials per SNR point.

\begin{figure}[t]
\centering
\includegraphics[width=3.4in]{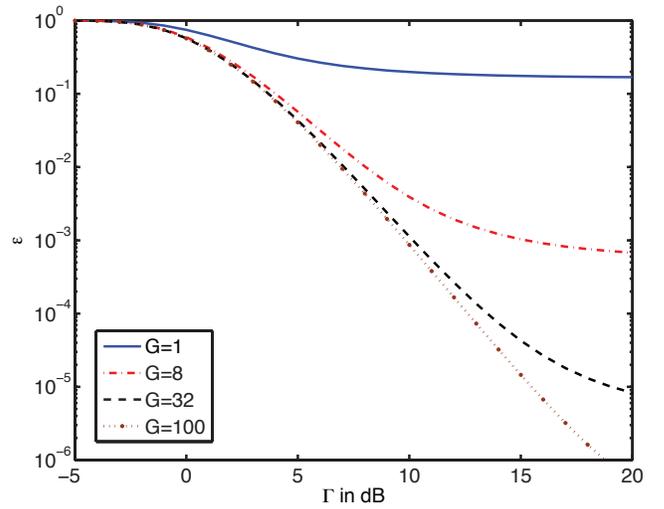}
\vspace{-0.2cm}
\caption{ Outage probability as a function of SNR $\Gamma$, conditioned on
the network shown in Fig. \ref{Figure:FigA} with the mixed-fading model. Performance is
shown both with and without spreading for several values of processing gain
($G$).}
\label{Figure:FigB}
\vspace{-0.4cm}
\end{figure}

\textbf{Example \#3.} The outage probabilities of examples \#1 and \#2 are
very high. One way to reduce the outage probability is to use spread
spectrum. Suppose that direct-sequence spread spectrum is used with a
processing gain of $G$ and $h=2/3$. In Fig. \ref{Figure:FigB}, the outage
probability is shown for direct-sequence networks using three different
processing gains as well as for an unspread network ($G=h=1$). A mixed fading
model ($m_{0}=4 $ and $m_{i}=1$ for $i\geq1$) is used. From this plot, a
dramatic reduction in outage probability when using direct-sequence
spreading can be observed.

\section{Spatial Averaging}

\label{Section:SpatialAveraging}

\begin{figure}[t]
\centering
\includegraphics[width=3.4in]{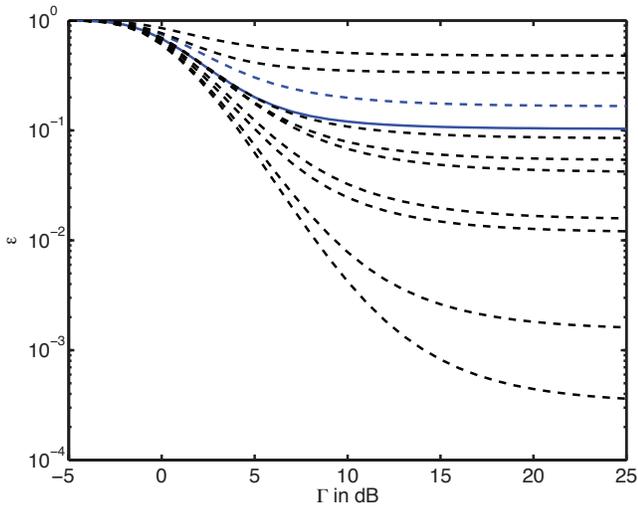}
\vspace{-0.2cm}
\caption{ Outage probability conditioned on several different network realizations. A uniform clustering
model is assumed with $r_{net}=1.0$, $r_{ex}=0.05$, $M=28$, mixed fading,
$\alpha = 3.5$, and no spreading. Ten network realizations were drawn, and the conditional
outage probabilities are indicated by dashed lines. The average outage
probability over 10,000 network realizations $\bar{\protect\epsilon}$ is
shown by the solid line. }
\label{Figure:FigC}
\vspace{-0.4cm}
\end{figure}

Because it is conditioned on ${\boldsymbol{\Omega}}$, the outage probability
will vary from one network realization to the next. The variability in
outage probability is illustrated in Fig. \ref{Figure:FigC}, which shows the
outage probability for ten different network realizations. One of the
networks is the one shown in Fig. \ref{Figure:FigA} while the other nine
networks were each realized in the same manner; i.e., with $M=28$
interferers drawn from a uniform clustering process with $r_{ex}=0.05$ and $%
r_{net}=1$ and the reference transmitter placed at distance $||X_{0}||=0.1$
from the reference receiver. The same set of parameters ($\alpha$, $\beta$, $%
p_{i}$, $P_{i}$, and $m_{i}$) used to generate the mixed-fading results of
Example \#2 were again used, and signals were not spread. From the figure,
it can be seen that the outage probabilities of different network
realizations can vary dramatically.

In addition to the location of the interferers, ${\boldsymbol{\Omega}}$
depends on the realization of the shadowing.   The shadowing factors $\{\xi_i\}$
can have any arbitrary distribution and need not be the same for all $i$.
In subsequent examples, log-normal shadowing is assumed, and the shadow 
factors are i.i.d. Gaussian with a standard deviation $\sigma _{s}$ and zero mean.

The conditioning on ${\boldsymbol{\Omega }}$
can be removed by marginalizing $\bar{F}_{\mathsf{Z}}(z|\boldsymbol{\Omega }%
) $ with respect to the spatial distribution of the network and the
distribution of the shadowing. This can be done analytically under certain
limitations. For instance, with the reference receiver at the center of the network, a
uniform spatial distribution, and no shadowing, the average outage
probability can be found, as shown in the Appendix. For more general cases
of interest, such as correlated spatial distributions (including uniform
clustering processes and Mat\'{e}rn  processes \cite{card}) or the presence
of shadowing, closed-form expressions are not readily available. However,
the outage probability can be estimated through Monte Carlo simulation by
generating a number of ${\boldsymbol{\Omega }}$, computing the outage
probability of each, and taking the average. Suppose that $N$ networks are
generated, and let ${\boldsymbol{\Omega =}}\boldsymbol{\Omega }_{i}$ for the 
$i^{th}$ network. Then the average outage probability may be found from the
Monte Carlo estimate of the complementary cdf: 
\begin{equation}
\bar{F}_{\mathsf{Z}}(z)=\frac{1}{N}\sum_{i=1}^{N}\bar{F}_{\mathsf{Z}}(z|%
\boldsymbol{\Omega }_{i}).
\end{equation}%
The average outage probability is 
\begin{equation}
\bar{\epsilon}=1-\bar{F}_{\mathsf{Z}}\left( \Gamma ^{-1}\right) .
\end{equation}%
As an example, the solid line in Fig. \ref{Figure:FigC} shows the
corresponding average outage probability averaged over $N=10,000$ network
realizations.

\begin{table}[ptb]
\caption{Average outage probability when the receiver is at the center of
the network for various values of $M$, $\protect\alpha$, and $G$ with
uniformly distributed interferers and no shadowing. The column marked $\bar{%
\protect\epsilon}_s$ is the average obtained through simulation while the
column marked $\bar{\protect\epsilon}_t$ is the theoretical value. }
\label{Table:Theory}\centering
\par
\begin{tabular}{|c|c|c|c|c|}
\hline
$M$ & $\alpha$ & $G$ & $\bar{\epsilon}_s$ & $\bar{\epsilon}_t$ \\ \hline
30 & 3 & 1 & $1.730 \times 10^{-1}$ & $1.711 \times 10^{-1}$ \\ \cline{3-5}
&  & 32 & $1.874 \times 10^{-3}$ & $1.906 \times 10^{-3}$ \\ \cline{2-5}
& 4 & 1 & $1.303 \times 10^{-1}$ & $1.290 \times 10^{-1}$ \\ \cline{3-5}
&  & 32 & $3.010 \times 10^{-3}$ & $3.118 \times 10^{-3}$ \\ \cline{1-5}
60 & 3 & 1 & $3.572 \times 10^{-1}$ & $3.624 \times 10^{-1}$ \\ \cline{3-5}
&  & 32 & $3.130 \times 10^{-3}$ & $3.229 \times 10^{-3}$ \\ \cline{2-5}
& 4 & 1 & $2.518 \times 10^{-1}$ & $2.574 \times 10^{-1}$ \\ \cline{3-5}
&  & 32 & $5.429 \times 10^{-3}$ & $5.671 \times 10^{-3}$ \\ \cline{1-5}
\end{tabular}%
\vspace{-0.3cm}
\end{table}

Table \ref{Table:Theory} compares the spatially averaged outage probability
obtained using the theoretical expression from the Appendix against the
corresponding value obtained by performing a Monte Carlo averaging over $%
N=10,000$ network realizations. For the results shown in Table \ref%
{Table:Theory}, the network radius is normalized to $r_{net}=1$, the
reference transmitter placed at distance $||X_{0}||=0.1$, the SINR threshold
set to $\beta =0$ dB, and the activity factor set to $p_{i}=0.5$. An
unshadowed mixed-fading channel was used with SNR set to $\Gamma =10$ dB. An
exclusion zone of radius $r_{ex}=0.05$ surrounds just the reference
receiver. Exclusion zones are not placed around the interferers or reference
transmitter, which ensures that the interferers are independently placed
according to a uniform distribution over the annulus of inner radius $r_{ex}$
and outer radius $r_{net}$. Table \ref{Table:Theory} shows the average
outage probability for $M=\{30,60\}$, $\alpha =\{3,4\}$, and $G=\{1,32\}$.
As can be seen, there is a good agreement between the theoretical values and
those obtained by simulating the interferer locations. The Monte Carlo
approach has an uncertainty that decreases with increasing $N$, and is an
effective approach to handle those cases that cannot be solved with the
analytical expression of the Appendix.

\begin{figure}[t]
\centering
\vspace{-0.2cm}
\includegraphics[width=3.4in]{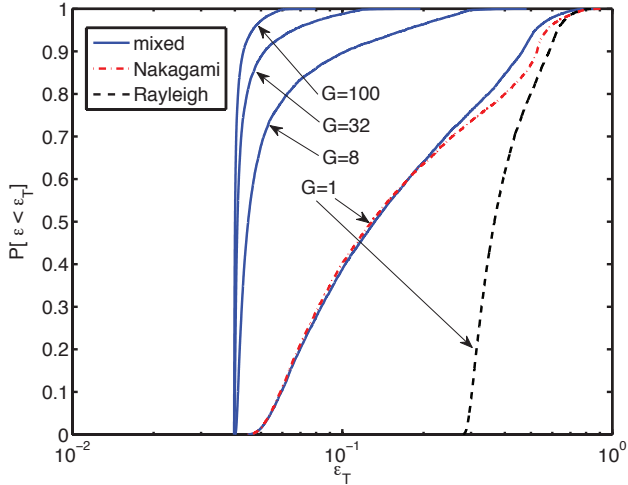}
\vspace{-0.2cm}
\caption{ Probability that the
conditional outage probability $\protect\epsilon $ is below the threshold
outage probability $\protect\epsilon _{_{T}}$ in a network with $%
r_{net}=1.0 $, $r_{ex}=0.05$, $M=28$, and $\Gamma =5$ dB. $N=10,000$ network
realizations were drawn to produce the figure. Results are shown for the
three fading models without spreading ($G=1$) and for the mixed-fading
model with spreading ($G=\{8,32,100\}$). }
\label{Figure:FigD}
\vspace{-0.3cm}
\end{figure}

Although the spatially averaged outage probability $\bar{\epsilon}$ provides
some insight into the average network behavior, it does not capture the
dynamics of the spatial distribution. A more useful metric for quantifying
the spatial variability is the probability that the conditional outage 
probability $\epsilon$ is either above
or below a threshold $\epsilon_{T}$. In particular $P[\epsilon>\epsilon_{T}]$
represents the fraction of network realizations that fail to meet a minimum
required outage probability at the reference receiver and can be construed as a \emph{network} outage
probability. The complement of the network outage probability $P[\epsilon
\leq\epsilon_{T}]$ is the cdf of $\epsilon$: $F_{\epsilon}(\epsilon_{T})$.
This probability is shown in Fig. \ref{Figure:FigD} for the three fading
models without spreading, and for the mixed-fading model with spreading. For
each curve, the SNR is $\Gamma$ = 10 dB and the system parameters are the
same as the corresponding Examples given in Section \ref{Section:Examples}. 
Each curve was generated by drawing $%
N=10,000$ networks using the same spatial model used to generate the
Examples; i.e., with $M=28$ interferers drawn from a uniform clustering
process with $r_{ex}=0.05$, $r_{net}=1$, and no shadowing. For each network,
the outage probability was computed and compared against the threshold $%
\epsilon_{T}$. The curves show the fraction of networks whose $\epsilon$
does not exceed the threshold.  Note that the curves become steeper with increasing $G$, 
which shows that spreading has the effect of making performance less sensitive
to the particular network topology.

\begin{figure}[t]
\centering
\includegraphics[width=3.4in]{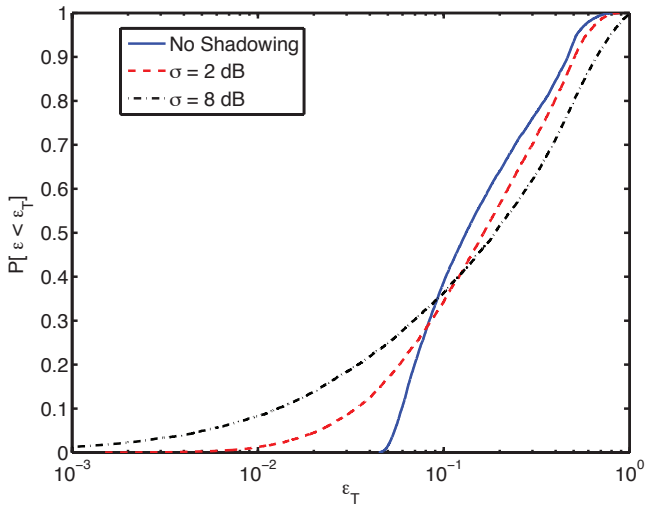}
\vspace{-0.2cm}
\caption{ Probability that the
conditional outage probability $\protect\epsilon $ is below the threshold
outage probability $\protect\epsilon _{_{T}}$ in a network with $r_{ex}=0.05$%
, $r_{net}=1.0$, $M=28$, $\Gamma =5$ dB, mixed fading, and no spreading ($%
G=1 $). $N=10,000$ network realizations were drawn to produce the figure.
Curves shown for no shadowing, and shadowing with two values of $\protect%
\sigma _{s} $. }
\label{Figure:FigE}
\vspace{-0.45cm}
\end{figure}

Shadowing can be incorporated into the model by simply drawing an
appropriate set of independent shadowing factors $\{\xi_{i}\}$ for each
network realization and using them to compute the $\{\Omega_{i}\}$ according
to (\ref{eqn:omega}). In Fig. \ref{Figure:FigE}, shadowing was applied to
the same set of $N=10,000$ networks used to generate Fig. \ref{Figure:FigD}.
Two standard deviations were considered for the log-normal shadowing: $%
\sigma _{s}=2$ dB and $\sigma_{s}=8$ dB, and again $\alpha = 3.5$. For each shadowed network
realization, the outage probability was computed for the mixed-fading model
without spreading ($G=1$). All other parameter values are the same ones used to
produce Fig. \ref{Figure:FigD}. The figure shows $P[\epsilon\leq%
\epsilon_{T}] $, and the no-shadowing case is again shown for reference. As
can be seen from the figure, the presence of shadowing and increases in $%
\sigma_{s}$ increase the variability of the conditional outage probability,
as indicated by the reduced slope of the cdf curves. Interestingly,
shadowing does not significantly alter the average outage probability in
this example. For low thresholds, such as $\epsilon_{T}<0.1$, performance is
actually better with shadowing than without. This behavior is presumably due
to the fact that, in shadowing, the reference signal power will sometimes be
much higher than without shadowing.

\begin{table}[ptb]
\caption{Average outage probability when the receiver is at the center ($%
\bar{\protect\epsilon}_{c}$) and on the perimeter ($\bar{\protect\epsilon}%
_{p}$) of the network for various $M$, $\protect\alpha$, $G$, and $\protect%
\sigma_{s}$. }
\label{Table:Position}\centering
\par
\begin{tabular}{|c|c|c|c||c|c||}
\hline
\multicolumn{4}{|c||}{Parameters} & \multicolumn{2}{|c||}{Outage
probabilities} \\ \hline
$M$ & $\alpha$ & $G$ & $\sigma_{s}$ & $\bar{\epsilon}_{c}$ & $\bar{\epsilon }%
_{p}$ \\ \hline
30 & 3 & 1 & 0 & 0.1528 & 0.0608 \\ \cline{4-6}
&  &  & 8 & 0.2102 & 0.0940 \\ \cline{3-6}
&  & 32 & 0 & 0.0017 & 0.0012 \\ \cline{4-6}
&  &  & 8 & 0.0112 & 0.0085 \\ \cline{2-6}
& 4 & 1 & 0 & 0.1113 & 0.0459 \\ \cline{4-6}
&  &  & 8 & 0.1410 & 0.0636 \\ \cline{3-6}
&  & 32 & 0 & 0.0028 & 0.0017 \\ \cline{4-6}
&  &  & 8 & 0.0123 & 0.0089 \\ \hline
60 & 3 & 1 & 0 & 0.3395 & 0.1328 \\ \cline{4-6}
&  &  & 8 & 0.4102 & 0.1892 \\ \cline{3-6}
&  & 32 & 0 & 0.0030 & 0.0017 \\ \cline{4-6}
&  &  & 8 & 0.0163 & 0.0107 \\ \cline{2-6}
& 4 & 1 & 0 & 0.2333 & 0.0954 \\ \cline{4-6}
&  &  & 8 & 0.2769 & 0.1247 \\ \cline{3-6}
&  & 32 & 0 & 0.0052 & 0.0027 \\ \cline{4-6}
&  &  & 8 & 0.0184 & 0.0117 \\ \hline
\end{tabular}%
\vspace{-0.45cm}
\end{table}

In a finite network, the average outage probability depends on the location
of the reference receiver. Rather than leaving the reference receiver at the
origin, Table \ref{Table:Position} explores the change in performance when
the reference receiver moves from the center of the radius-$r_{net}$
circular network to the perimeter of the network.  When the reference receiver
is at the perimeter of the network, the circular network region is centered
at coordinate $-r_{net}$. The SNR was
set to $\Gamma=10$ dB, a mixed-fading channel model was assumed, and 
other parameter values are the same ones used to produce Fig. \ref%
{Figure:FigE}; i.e., $r_{ex}=0.05$, $r_{net}=1$, $\beta=0$ dB, and $%
p_{i}=0.5 $. The interferers were placed according to the uniform clustering
model and the reference transmitter was placed at distance $||X_{0}||=0.1$
from the reference receiver. For each set of values of the parameters $G$, $%
\alpha,$ $\sigma_{s},$ and $M$, the outage probability at the network center 
$\bar{\epsilon}_{c}$ and at the network perimeter $\bar{\epsilon}_{p}$ were
computed by averaging over $N=10,000$ realizations of mobile placement and
shadowing. Two values of each parameter were considered: $G=\{1,32\},$ $%
\alpha=\{3,4\},$ $\sigma_{s}=\{0,8\},$ $M=\{30,60\}.$ The table indicates
that $\bar{\epsilon}_{p}$ is considerably less than $\bar{\epsilon}_{c}$ in
the finite network, which cannot be predicted by the traditional
infinite-network model. This reduction in outage probability is more
significant for the unspread network, and is less pronounced with increasing 
$G$. Both $\bar{\epsilon}_{p}$ and $\bar{\epsilon }_{c}$ increase as $M$ and 
$\sigma_{s}$ increase and $G$ decreases.

As $\alpha$ increases, both $\bar{\epsilon}_{p}$ and $\bar{\epsilon}_{c}$
increase when $G=32$, but decrease when $G=1$. This difference occurs
because spread-spectrum systems are less susceptible to the near-far problem
than unspread ones. The increase in $\alpha$ is not enough to cause a
significant increase in the already high outage probability for unspread
systems in those realizations with interferers close enough to the reference
receiver to cause a near-far problem. In the same realizations, the less
susceptible spread-spectrum systems do experience a significantly increased
outage probability.

\section{Number of Interferers and Transmission Capacity}

\label{Transmission Capacity}

\begin{figure}[t]
\centering
\includegraphics[width=3.4in]{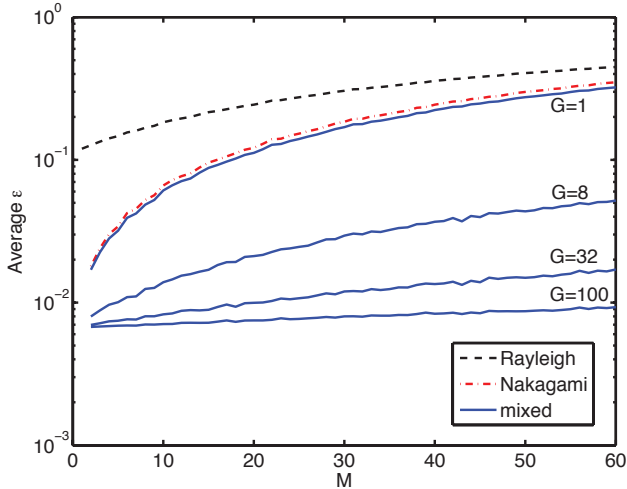}
\vspace{-0.25cm}
\caption{ Average outage
probability as a function of $M$ for SNR $\Gamma =10$ dB and log-normal
shadowing with $\protect\sigma _{s}=8$ dB. Results are shown for the three
fading models without spreading ($G=1$) and for the mixed-fading model with
spreading ($G=\{8,32,100\}$). }
\label{Figure:FigF}
\vspace{-0.4cm}
\end{figure}

Fig. \ref{Figure:FigF} shows the outage probability averaged over $10,000$
networks as a function of the number of interferers $M$ in the presence of
log-normal shadowing with standard-deviation $\sigma_{s}=8$ dB. A
uniform-clustering network model is again assumed with $r_{ex}=0.05$ and $%
r_{net}=1$, and the reference receiver is at the center of the network ($r=0$%
). The SNR is $\Gamma=10$ dB and the parameters $\alpha=3.5$, $p_{i}=0.5$,
and $\beta=0$ dB are again used. Results are shown for the unspread network
and the three channel models of Examples \#1 and \#2, as well as for the
mixed-fading channel model and three spreading factors $G=\{8,32,100\}$.
From the curves, it can be observed that performance improves with
increasing $G$, and that higher values of $G$ reduce the sensitivity to an
increased density of interfering mobiles. Both of these effects are
attributable to the improved interference suppression as the processing gain
increases.

A trade-off can be seen in Fig. \ref{Figure:FigF}. As $M$ increases, the
performance of each link degrades, yet there will be more simultaneous
transmissions supported by the network. This trade-off can be quantified by
the \emph{transmission capacity}, which represents the network throughput
per unit area \cite{weber}. It can be found by multiplying network density $%
\lambda$, which is the number of mobiles per unit area, by the rate that
bits are successfully transmitted over the reference link: 
\begin{equation}
\tau=\lambda(1-\epsilon)b
\end{equation}
where $b$ is the per-link throughput in the absence of an outage, in units
of bps, and the factor $(1-\epsilon)$ ensures that only successful
transmissions are counted.

\begin{figure}[t]
\centering
\includegraphics[width=3.4in]{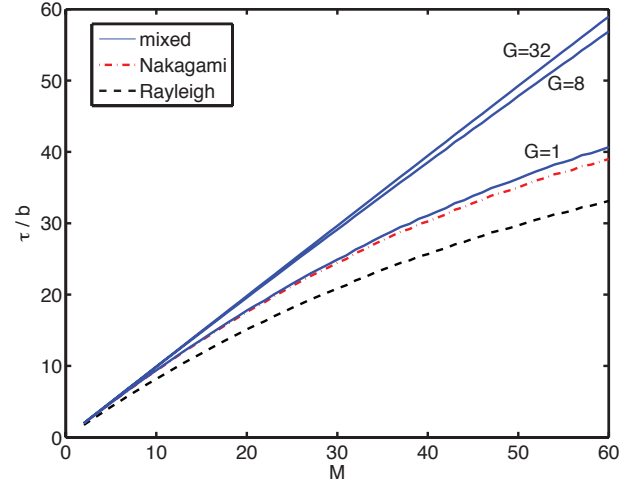}
\vspace{-0.25cm}
\caption{ Normalized
transmission capacity as a function of $M$ for SNR $\Gamma =10$ dB and
log-normal shadowing with $\protect\sigma _{s}=8$ dB. Results are shown for
the three fading models without spreading ($G=1$) and for the mixed-fading
model with spreading ($G=\{8,32\})$. The performance with $G=100$ is not
shown and is slightly higher than it is for $G=32$. }
\label{Figure:FigG}
\vspace{-0.4cm}
\end{figure}

Fig. \ref{Figure:FigG} shows the transmission capacity normalized by $b$ as
a function of $M$ for the same cases shown in Fig. \ref{Figure:FigF}. As can
be seen from the figure, spreading improves the transmission capacity due to
the corresponding reduction in outage probability. The curves become steeper
with increasing $G$ reflecting the superior ability of spreading to cope
with increasing network density, provided that enough bandwidth is available
to support the spread-spectrum signaling.

\setcounter{equation}{38}
\begin{figure*}[t]
\begin{eqnarray}
\bar{F}_{z}(z)
\hspace{-0.25cm}
& = & 
\hspace{-0.25cm}
e^{-\beta _{0}z}
\hspace{-0.15cm}
\sum_{s=0}^{m_{0}-1}{\left( \beta
_{0}z\right) }^{s}\sum_{t=0}^{s}\frac{z^{-t}}{(s-t)!}  
\hspace{-0.35cm}
\mathop{ \sum_{\ell_i \geq 0}}_{\sum_{i=0}^{M}\ell _{i}=t}
\hspace{-0.15cm}
\prod_{i=1}^{M}%
\left[ (1-p_{i})\delta _{\ell _{i}}+\frac{2p_{i}\Gamma (\ell _{i}+m_{i})}{%
\alpha c_{i}^{2/\alpha }(r_{net}^{2}-r_{ex}^{2})(\ell _{i}!)\Gamma (m_{i})}%
\displaystyle\int_{c_{i}r_{ex}^{\alpha }}^{c_{i}r_{net}^{\alpha }}
\hspace{-0.15cm}
\frac{%
x^{\left( \frac{2-\alpha }{\alpha }\right) }\left( \frac{1}{m_{i}x}\right)
^{\ell _{i}}}{\left( \frac{\beta _{0}}{m_{i}x}+1\right) ^{(m_{i}+\ell _{i})}}%
\;dx\right]  \nonumber \\
\vspace{-1cm}
\label{cdfwithintegral}
\end{eqnarray}
\vspace{0cm}
\hrulefill
\end{figure*}


\setcounter{equation}{43}
\begin{figure*}[t!]
\begin{eqnarray}
\bar{F}_{z}(z) 
&=&
\hspace{-0.25cm}
e^{-\beta _{0}z}
\hspace{-0.15cm}
\sum_{s=0}^{m_{0}-1}{\left( \beta
_{0}z\right) }^{s}\sum_{t=0}^{s}\frac{z^{-t}}{(s-t)!}  
\hspace{-0.35cm}
\mathop{ \sum_{\ell_i \geq 0}}_{\sum_{i=0}^{M}\ell _{i}=t}
\hspace{-0.15cm}
\prod_{i=1}^{M}%
\left[  (1-p_{i})\delta _{\ell _{i}}+\frac{2p_{i}\Gamma
(\ell _{i}+m_{i})m_{i}^{m_{i}}\left[
J\left( c_{i}r_{net}^{\alpha }\right) -J\left( c_{i}r_{ex}^{\alpha }\right) %
\right]}{\alpha
c_{i}^{2/\alpha }(r_{net}^{2}-r_{ex}^{2})(\ell _{i}!)\Gamma (m_{i})\beta
_{0}^{(m_{i}+\ell _{i})}\left( m_{i}+\frac{2}{\alpha }\right) } \right]  
\nonumber \\
\label{cdf}
\end{eqnarray}
\vspace{0cm}
\hrulefill
\end{figure*}
\setcounter{equation}{33}

\section{Conclusion}

\label{Section:Conclusions}

A new and potent method for evaluating the performance of an ad hoc network
has been presented. Since our model allows realistic deployments, it can be incorporated into time-dependent random geometric graphs that capture the dynamic network status.
Factors modeled include a finite spatial extent and
number of mobiles, any spatial distribution of the mobiles, the existence of
exclusion zones, path-loss, Nakagami fading, and shadowing. The Nakagami
m-parameter can vary among the mobiles provided that it is an integer for
the reference transmitter. Using the analysis, many aspects of the network
performance are illuminated. For example, one can determine the influence
of the choice of the spreading factor, the effect of the receiver location within the finite
network region, and the impact of the attenuation power law.

\appendix

\noindent By averaging over $\boldsymbol{\Omega }_{1}=\{\Omega
_{1},...,\Omega _{M}\}$, the complementary cdf of $\mathsf{Z}$ is 
\begin{equation}
\bar{F}_{\mathsf{Z}}(z)=\int f_{\boldsymbol{\Omega }}(\boldsymbol{\omega })%
\bar{F}_{\mathsf{Z}}(z\big|\boldsymbol{\omega })d\boldsymbol{\omega }
\label{cdf_M}
\end{equation}%
where $f_{\boldsymbol{\Omega }}(\boldsymbol{\omega })$ is the joint pdf of
$\boldsymbol{\Omega }_{1}$. Assuming that the $\{\Omega _{i}\}$ are
independent, 
\begin{equation}
f_{\boldsymbol{\Omega }}(\boldsymbol{\omega })=\prod_{i=1}^{M}f_{\Omega
_{i}}(\omega _{i})  \label{independent_pdf}
\end{equation}%
where $f_{\Omega _{i}}(\omega _{i})$ is the pdf of $\Omega _{i}$. Define $%
\mho _{i}=\Omega _{i}^{-1},i\geq 1,$ to be the \emph{inverse} normalized
power of the $i^{th}$ interferer. Using (\ref{independent_pdf}) and the
definition of $\mho _{i}$, (\ref{cdf_M}) may be expressed as 
\begin{equation}
\bar{F}_{\mathsf{Z}}(z)=\int \left( \prod_{i=1}^{M}f_{\mho
_{i}}(x_{i})\right) \bar{F}_{\mathsf{Z}}(z\big|\{x_{1}^{-1},..x_{M}^{-1}\})d%
\boldsymbol{x.}  \label{cdf_M_rho}
\end{equation}%
If the location of interferer $X_{i}$ is uniform within the annulus of inner
radius $r_{ex}$ and outer radius $r_{net}$, then its distance $%
r_{i}=||X_{i}||$ from the origin has pdf 
\begin{equation}
f_{r_{i}}(r)=%
\begin{cases}
\displaystyle\frac{2r}{r_{net}^{2}-r_{ex}^{2}} & 
\mbox{for $r_{ex} \leq r
\leq r_{net}$} \\ 
0 & \mbox{elsewhere.}%
\end{cases}%
\end{equation}%
From the definition of $\mho _{i}$ and (\ref{eqn:omega}), $\mho
_{i}=(G/h)(P_{0}/P_{i})r_{i}^{\alpha },i\geq 1,$ in the absence of
shadowing, and it follows that the pdf of $\mho _{i}$ is 
\begin{equation}
f_{\mho _{i}}(x)=%
\begin{cases}
\displaystyle\frac{2x^{\left( \frac{2-\alpha }{\alpha }\right) }}{\alpha
c_{i}^{2/\alpha }(r_{net}^{2}-r_{ex}^{2})} & 
\mbox{for $c_i r_{ex}^{\alpha}
\leq x \leq c_i r_{net}^{\alpha}$} \\ 
0 & \mbox{elsewhere}%
\end{cases}
\label{pdf_mho}
\end{equation}%
where $c_{i}=(G/h)(P_{0}/P_{i})$. Substituting (\ref%
{Equation:NakagamiConditional}) and (\ref{pdf_mho}) into (\ref{cdf_M_rho})
and extracting factors from the integral that are independent of the
variable of integration yields equation (\ref{cdfwithintegral}) at the top of the page, 
where $\beta _{0}=\beta m_{0}||X_{0}||^{\alpha }$ in the absence of fading.

\setcounter{equation}{39}

By factoring the denominator and splitting the integration interval, the
integral in (\ref{cdfwithintegral}) is 
\begin{equation}
\int_a^b
\frac{x^{\left( \frac{2-\alpha }{\alpha }\right) }\left( \frac{1}{m_{i}x}\right)
^{\ell _{i}}}{\left( \frac{\beta _{0}}{m_{i}x}+1\right) ^{(m_{i}+\ell _{i})}}%
\;dx=\frac{m_{i}^{m_{i}}}{\beta _{0}^{(m_{i}+\ell _{i})}}\left[ I\left(
b \right) -I\left( a\right) \right]
\label{split}
\end{equation}%
where $a= c_{i}r_{ex}^{\alpha }$, $b = c_{i}r_{net}^{\alpha }$, and 
\begin{equation}
I(y)=\int_{0}^{y}x^{\left( m_{i}+\frac{2}{\alpha }-1\right) }\left( 1+\frac{%
m_{i}x}{\beta _{0}}\right) ^{-(m_{i}+\ell _{i})}\;dx.
\end{equation}%
By performing the change of variable $x=y\nu $, 
\begin{align}
I(y)& =\int_{0}^{1}(y\nu )^{\left( m_{i}+\frac{2}{\alpha }-1\right) }\left(
1+\frac{m_{i}y\nu }{\beta _{0}}\right) ^{-(m_{i}+\ell _{i})}\;yd\nu  \notag
\\
& =y^{(m_{i}+\frac{2}{\alpha })}\int_{0}^{1}\nu ^{\left( m_{i}+\frac{2}{%
\alpha }-1\right) }\left( 1+\frac{m_{i}y}{\beta _{0}}\nu \right)
^{-(m_{i}+\ell _{i})}\;d\nu  \notag \\
& 
\hspace{-0.75cm}
=\frac{y^{(m_{i}+\frac{2}{\alpha })}}{\left( m_{i}+\frac{2}{\alpha }%
\right) }{_{2}F_{1}\left( \left[ m_{i}+\ell _{i},m_{i}+\frac{2}{\alpha }%
\right] ;m_{i}+\frac{2}{\alpha }+1;-\frac{m_{i}y}{\beta _{0}}\right) }
\label{Ihyp}
\end{align}%
where $_{2}F_{1}$ is  the Gauss hypergeometric function,
which has the integral representation: 
\begin{eqnarray}
_{2}F_{1}([a,b];c;x) & = & \nonumber \\
&  & \hspace{-2cm}
\frac{\Gamma (c)}{\Gamma (b)\Gamma (c-b)}%
\int_{0}^{1}\nu ^{b-1}(1-\nu )^{c-b-1}(1-x\nu )^{-a}d\nu  \nonumber \label{HYPERG} \\
\end{eqnarray}%
and $c=b+1$, $\Gamma (c)/(\Gamma (b)\Gamma (c-b))=b$. The hypergeometric
function can be represented by an infinite series if $|x|<1$ and is
evaluated by analytical continuation if $|x|\geq 1$ \cite{arfken,pearson}.
The function is widely known and is implemented as a single function call in
most mathematical programming languages, including Matlab. The complementary
cdf of $\mathsf{Z}$ can be found by substituting (\ref{split}) into (\ref%
{cdfwithintegral}), with the $I(\cdot )$ function as defined in (\ref{Ihyp}%
). The result is equation (\ref{cdf}) at the top of the page, where
\setcounter{equation}{44}
\begin{equation}
J(y)= 
{_{2}F_{1}\left( \left[ m_{i}\hspace{-0.05cm}+\hspace{-0.05cm}\ell _{i},m_{i}\hspace{-0.05cm}+\hspace{-0.05cm}\frac{2}{\alpha }\right]
;m_{i}\hspace{-0.05cm}+\hspace{-0.05cm}\frac{2}{\alpha }\hspace{-0.05cm}+\hspace{-0.05cm}1;-\frac{m_{i}y}{\beta _{0}}\right)}
y^{m_i+\frac{2}{\alpha}}
\end{equation}

\vfill
\pagebreak

\section*{Acknowledgements}
The authors would like to thank Salvatore Talarico for his programming assistance and suggestions related to the Appendix.

\bibliographystyle{ieeetr}

\ifpdf
  \begin{IEEEbiography}{Don Torrieri}
\else
  \begin{IEEEbiography}[{\includegraphics[width=1in,height=1.25in,clip,keepaspectratio]{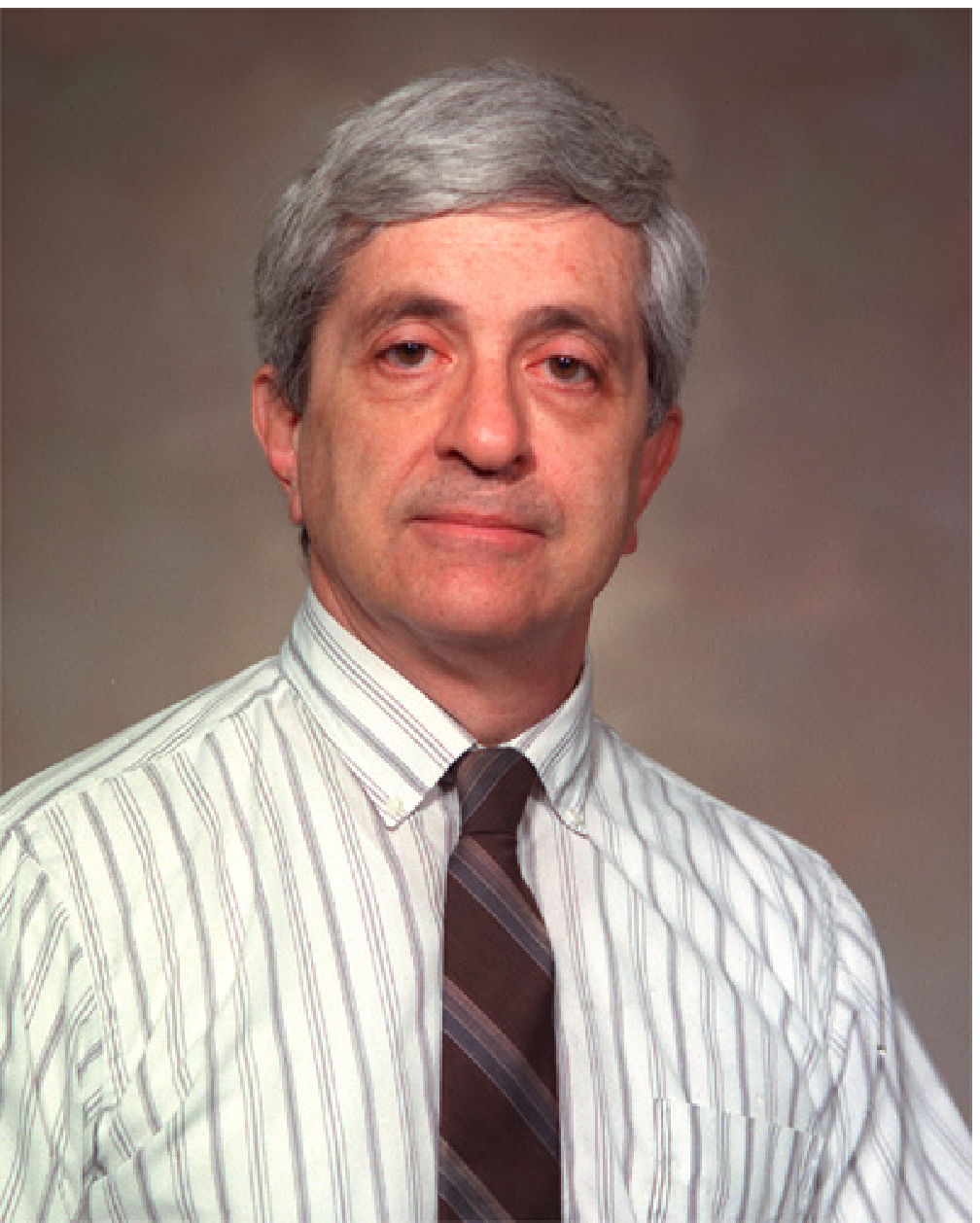}}]{Don Torrieri}
\fi
 is a research engineer and Fellow of the US Army Research Laboratory. His primary research interests are communication systems, adaptive arrays, and signal processing. He received the Ph. D. degree from the University of Maryland. He is the author of many articles and several books including {\em Principles of Spread-Spectrum Communication Systems}, 2nd ed. (Springer, 2011). He teaches graduate courses at Johns Hopkins University and has taught many short courses. In 2004, he received the Military Communications Conference achievement award for sustained contributions to the field.
\end{IEEEbiography}

\ifpdf
  \begin{IEEEbiography}{Matthew C. Valenti}
\else
  \begin{IEEEbiography}[{\includegraphics[width=1in,height=1.25in,clip,keepaspectratio]{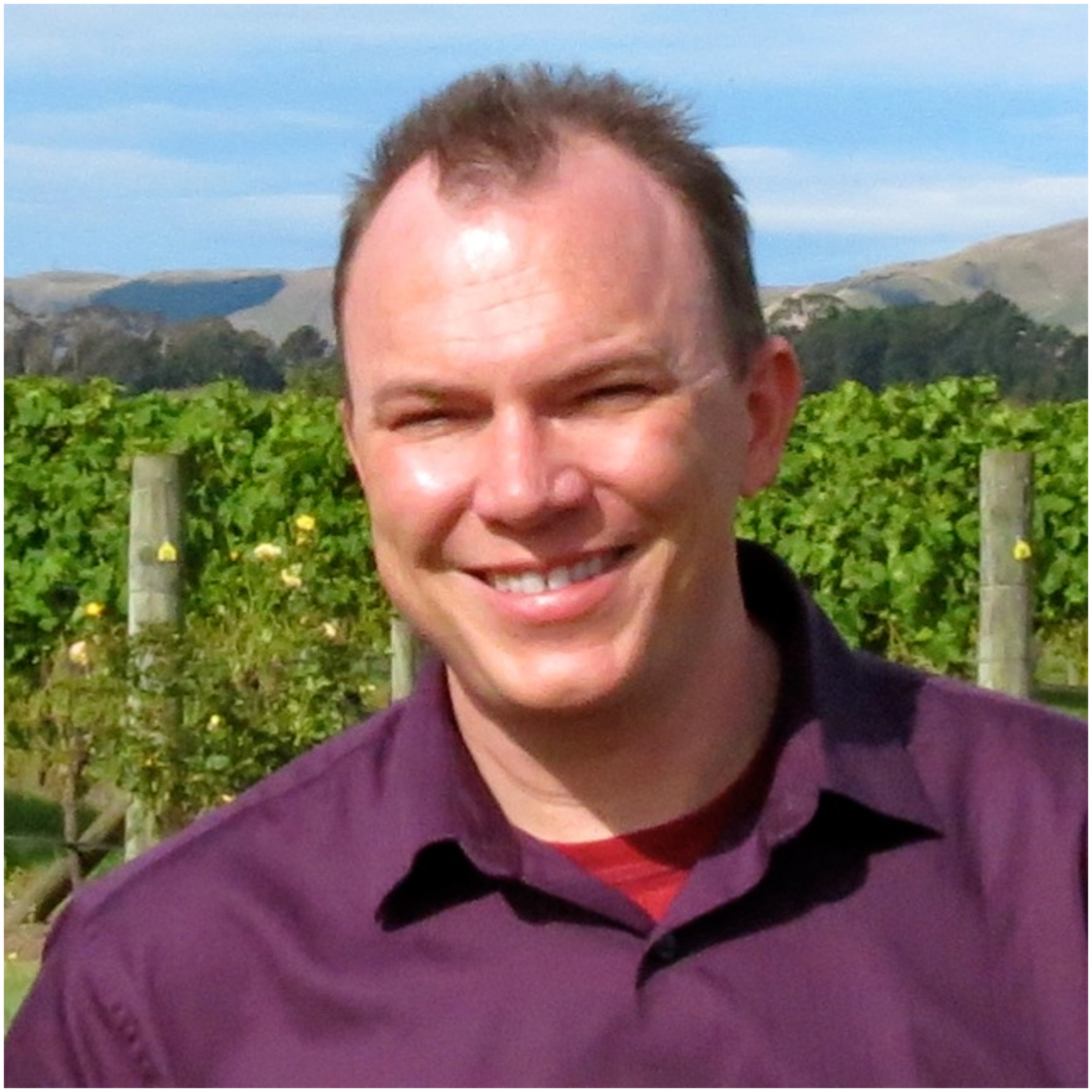}}]{Matthew C. Valenti}
\fi
is a Professor in Lane Department of Computer Science and Electrical Engineering at West Virginia University. He holds BS and Ph.D. degrees in Electrical Engineering from Virginia Tech and a MS in Electrical Engineering from the Johns Hopkins University. From 1992 to 1995 he was an electronics engineer at the US Naval Research Laboratory.  He serves as an associate editor for {\em IEEE Wireless Communications Letters} and as Vice Chair of the Technical Program Committee for Globecom-2013.  Previously, he has served as a track or symposium co-chair for VTC-Fall-2007, ICC-2009, Milcom-2010, ICC-2011, and Milcom-2012, and has served as an editor for {\em IEEE Transactions on Wireless Communications} and {\em IEEE Transactions on Vehicular Technology}. His research interests are in the areas of communication theory, error correction coding, applied information theory, wireless networks, simulation, and secure high-performance computing.  His research is funded by the NSF and DoD.  He is registered as a Professional Engineer in the State of West Virginia.
\end{IEEEbiography}

\vfill

\end{document}